# SECURITY ISSUES IN SPEECH WATERMARKING FOR INFORMATION TRANSMISSION

Rupa Patel , Urmila Shrawankar

*Department of Computer Science & Engineering*
*G. H. Raisoni College of Engineering, Nagpur*

## ABSTRACT

*The secure transmission of speech information is a significant issue faced by many security professionals and individuals. By applying voice-encryption technique any kind of encrypted sensitive speech data such as password can be transmitted. But this has the serious disadvantage that by means of cryptanalysis attack encrypted data can be compromised. Increasing the strength of encryption/decryption results in an associated increased in the cost. Additional techniques like stenography and digital watermarking can be used to conceal information in an undetectable way in audio data. However this watermarked audio data has to be send through unreliable media and an eavesdropper might get hold of secret message and can also determine the identity of a speaker who is sending the information since human voice contains information based on its characteristics such as frequency, pitch, and energy. This paper proposes Normalized Speech Watermarking technique. Speech signal is normalized to hide the identity of the speaker who is sending the information and then speech watermarking technique is applied on this normalized signal that contains the message (password) so that what information is transmitted should not be unauthorizedly revealed.*

**KEYWORDS:** Speech Normalization, Speech Watermarking, Secure Transmission , Voice Activity Detection,

## 1. INTRODUCTION

With the increase in mobile and internet communication speech signals are often used for information transmission. Speech watermarking technique is generally used to conceal the secret messages .However this secret audio data is transmitted using untrusted , unreliable media and an eavesdropper might realize that secret communication has taken place. Using logarithmic-exponential watermark method the security level of watermarking could be improved[52]. But ,if eavesdropper might get hold of this secret message  he would determine the identity of the speaker who is sending the information .Since Yin Yin suggested that from the voice of speakers the identity of the speaker can be determined [27] and human voice contains voice information based on its characteristics such as frequency, pitch, and energy. The security of present watermarking technique can be further enhanced by hiding the speaker identity using additional normalization technique. In this paper we aim at  devising effective and robust speaker independent method  that will be  robust against speech variations to hide speaker identity and  concealing the speech secret information for transmission so that third party don't even realize that the communication has taken place.

   This paper is organized into five sections. In the next section we describe system architecture, techniques are reviewed in section 3, implementation steps of proposed algorithm are given in section 4, experimental results are discussed in section 5, followed by conclusion and discussion.

## 2. SYSTEM ARCHITECTURE

As shown in figure1 our proposed system contains two main components

   A. Speech Normalization

   This module deals with the normalization of the voice speech to hide speaker identity.  It contains two sub modules
1. Classification of Voiced/Unvoiced signals

The input to this will be speech signal. Using voice activity detector signal will be classified as voiced or unvoiced.

2. Voice Normalization
   In this module filter voiced speech will be normalized in specific frequency range.
B. Speech Watermarking
   This module deals with concealing the password so that integrity of information could be maintained.
   .

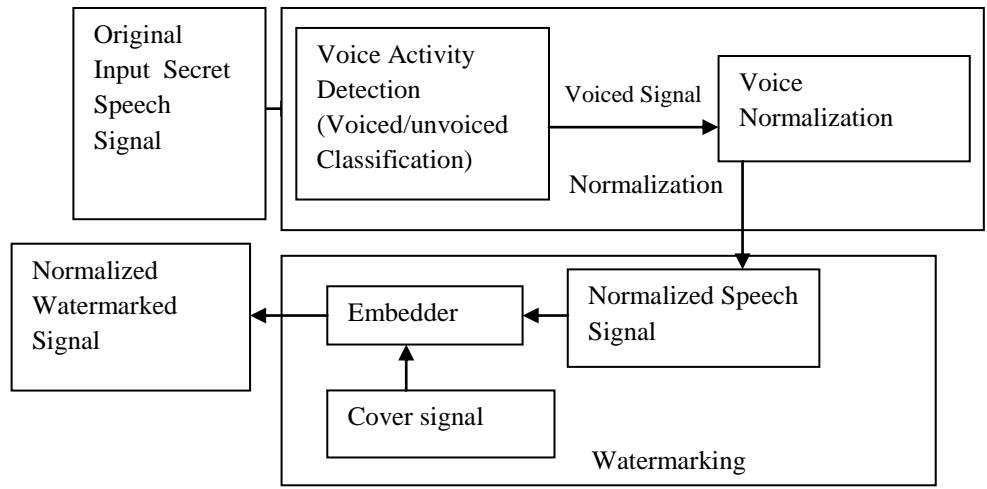

Figure 1: Proposed system architecture

# 3. METHODOLOGY

## 3.1 Voice Activity Detection

The presence of speech in the input audio signal can be detected by using Voice Activity Detection techniques. The decision whether audio is voiced-unvoiced is based on the different features extracted from speech signals. The various voice activity detection techniques are Energy, Zero-Crossing Rate, AMDF, spectral entropy, Teager Energy operator. In the proposed work ZCR and short time energy powerful techniques that works in time domain for speech activity detection have been applied.

### 3.1.1 Short Time Energy

Since the speech signals is not stationary and its characteristics changes with time signals are divided into short sections called frames. The energy of the signal also changes with time. The short-term energy $E_n$ of the voice signal x(m) ,window function w(n) *is* defined as follows [50].

$$E_n = \sum_{m=-\infty}^{\infty} [x(m)w(n-m)]^2$$

### 3.1.2 Zero crossing Rate

Considering audio data as discrete signals Zero crossing is said to occur if successive samples have different algebraic signs. The rate at which zero crossings occur is a simple measure of the frequency content of a signal. The short time zero rate $Z_n$ of the sampled speech signal is described as [50]

$$Z_n = \sum_{m=-\infty}^{\infty} |\text{sgn}[x(m)] - \text{sgn}[x(m-1)]| w(n-m)$$

where

$$\text{sgn}[x(n)] = \begin{cases} 1, & x(n) \geq 0 \\ -1, & x(n) < 0 \end{cases}$$

and

$$w(n) = \begin{cases} \dfrac{1}{2N} & for, 0 \leq n \leq N-1 \\ 0 & for, otherwise \end{cases}$$

Generally voiced signal has high energy and low ZCR while unvoiced signal has low energy and high ZCR.

## 3.2   PITCH DETECTION TECHNIQUE

The pitch information mainly lies on the voiced part in the speech signal. The fundamental frequency (*F0*) is the main cue of the pitch. The algorithms that are available for estimating pitch are autocorrelation, cepstral method ,simplified inverse filtering techniques, spectral linear predictive coding and AMDF( average magnitude difference function)[40].Cepstral method has been used in this proposed work and its detail working is given below.

### 3.2.1    Cepstral Method

The cepstrum, defined as the real part of the inverse Fourier transform of the log-power spectrum, has a strong peak corresponding to the pitch period of the voiced speech segment being analyzed.[52,53]. The cepstral coefficients are found by using the following equation

$$C(\tau) = \left| F\left\{ \log \left| F(x(t)) \right|^2 \right\} \right|^2$$

$$C(\tau) = F^{-1}\left\{ \log \left| F(x(t)) \right|^2 \right\}$$

where $F$ denotes fourier transform and $x[n]$ is the signal in the time domain. $|F\{x[n]\}|2$ is the
power spectrum estimate of the signal. t is the quefrency. The fundamental frequency is estimated by can be obtained using

$$\hat{f}_1 = \frac{1}{\tau_{max}}, \; C(\tau_{max}) = \max_{\tau} C(\tau)$$
$$\tau > 0$$

## 3.3   SPEECH NORMALIZATION

The Fundamental variations in speech exist both between speakers and within the speech of a single speaker. The variability in the acoustic speech signal is due to the factors such as noise, varying channel conditions, varying speaking rates and speaker specific differences [1]. ]. A listener despite of this variability, is able to identify and understand speech spoken by different speakers .The aim of speech normalization is to reduce the speaker variability by modifying the spectral (or Cepstral) representation of a incoming speech waveforms so that the    words (message /phoneme) uttered by   speakers will   be transformed into a speaker-invariant representation. The different normalization techniques are described below

### 3.3.1    Histogram Equalization

Histogram Equalization [44] proposes generalizing the normalization to all the statistical moments by transforming the Cepstral coefficients probability density function. Histogram equalization **n**ormalize histogram of the speech features , but it does not consider information  in the acoustic model[47].The histogram equalization method assumes that the transformation is monotonic and does not cause an information loss. In the case of speech analysis, the speech signal is segmented into frames and each frame is represented by a feature vector. Method that is generally is apply the histogram equalization to each component of the feature vector representing each frame of the speech signal. In order to obtain the transformation for each component, the cumulative histogram is to been estimated. The transformation is then computed for the points in the center of each interval and then could be  applied to the parameters to be compensated as a linear interpolation using the closest couple of points for which the transformation was computed[11,12].

### 3.3.2        Cepstral Mean and variance

Statistical Matching Algorithms define linear and non linear transforms in order to modify the noisy features statistics and make them equal to those of a reference set of clean data. The most relevant algorithm is  Cepstral Mean ad Variance Normalization. The additive effect of noise implies a shift on the average of the MFCC coefficients probability density function added to a scaling of its variance. Given a noisy Cepstral coefficient $y$ contaminated with an additive noise with mean value $h$, and given the clean Cepstral coefficient $x$ with mean value $\mu x$ and variance $\sigma x$, the contaminated MFCC $y$ , representing  the variance scaling produced

$$y = \alpha \cdot x + h$$

$$\mu_y = \alpha \cdot \mu_x + h$$

$$\sigma_y = \alpha \cdot \sigma_x$$

If we normalize the mean and variance of both coefficients $x$ and $y$, their expressions

will be

$$\hat{x} = \frac{x - \mu_x}{\sigma_x}$$

$$\hat{y} = \frac{y - \mu_y}{\sigma_y} = \frac{(\alpha \cdot x + h) - (\alpha \cdot \mu_x + h)}{\alpha \cdot \sigma_x} = \hat{x}$$

The above equation shows that CMVN makes the coefficients robust against the shift and scaling introduced by noise. The linear transformation performed by CMNV only eliminates the linear effects of noise. The non-linear distortion produced by noise does not only affect the mean and variance of the probability density functions but it also affects the higher order moments [51].

### 3.3.3        Vocal tract Normalization

Vocal tract length normalization (VTLN) is an acoustic normalization technique designed to decrease speech variability due to differing vocal tract lengths among speakers This can be compensated for by stretching or compressing the frequency axis of the spectrum prior to recognition. If VTLN is applied, the stretching/compressing is applied during the feature extraction process, after the windowing and the Fourier transform. In the VTLN algorithm, the frequency axis is stretched or compressed, this is also called scaling or warping i.e. it requires a transformation in the frequency domain known as frequency warping .There are two main methods .Formant-based normalization that determines the warping factor directly based on the location of the formant frequencies and Maximum-likelihood-based normalization that generates a model and designates the warping factor that maximizes the likelihood of the test data given the statistical model. [53].For speech normalization we will perform a transformation in the frequency domain.

## 3.4 SPEECH WATERMARKING

Digital speech Watermarking seeks to hide information inside another audio , but should be resilient to intentional or unintentional manipulations and resistant to watermark attacks .The embedded data should also maintain the quality of the host/cover signal, According to Human Perception, the watermarking techniques can be divided into three types Visible Watermark, Invisible Watermark, Dual Watermark. According to Working Domain, the watermarking techniques can be divided into two types Time Domain Watermarking techniques and Frequency Domain Watermarking techniques. According to the watermarking extraction process, techniques can be divided into three types Non-blind, Semi-blind, Blind. For the proposed idea the invisible, blind and frequency masking approach will be used to embedded the normalized watermark signal with the musical cover signal. Blind watermarking does not use the watermark during extraction process and is superior over other watermarking involving watermark for extraction. Invisible watermark method is embedding watermark into the data in such a way that the changes made to the signal components are perceptually not noticed and generally frequency domain methods are more robust than time domain techniques[18,52].

## 4. PROPOSED ALGORITHM

The Steps that will be performed for implementing the proposed Normalized Speech Watermarking technique are

1. Voice activity detection (identification of voiced /unvoiced speech signals)
   a. Input audio speech waveform
   b. For voiced/unvoiced/silence classification compute short time energy ( E ) and Zero crossing Rate (ZCR) using Frame by Frame processing.
   c. If short time energy is high and ZCR is low then speech waveform is voiced signal otherwise unvoiced.

Figure 2 illustrates above steps.

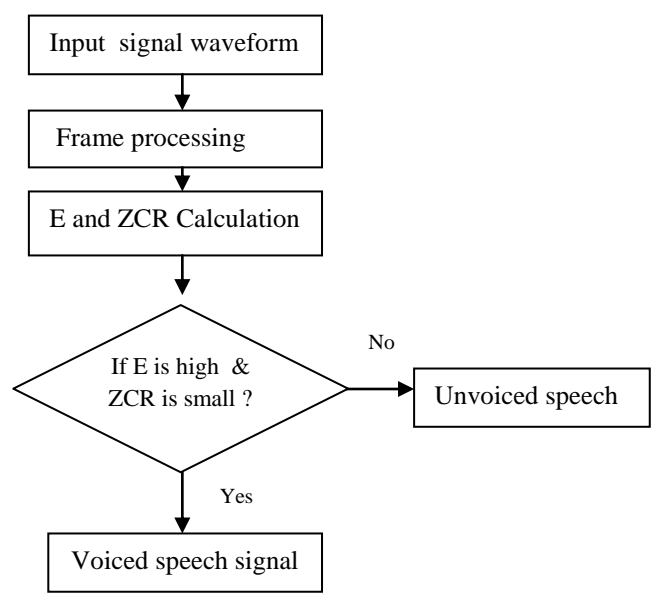

Figure 2: Classification of voiced/unvoiced speech signal

2. Normalization of voiced signal.
   a. Apply noise removal filter to voiced speech signal
   b. Estimate Pitch of the voiced speech signal
      i. Compute cepstral coefficients
      ii. Estimate cepstral peak
   c. Normalize the voiced speech signal within specific range

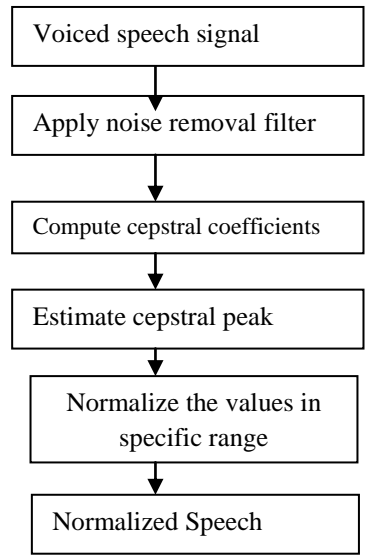

Figure 3: Voice normalization

3. Apply Speech watermarking to normalized speech signal
   a. Apply logarithmic operation to normalized speech that will generate normalized watermark.
   b. Input Musical cover signal
   c. Find the center density of high frequency subband of cover signal.
   d. Compute the length of both cover and watermark signal.
   e. If the length of normalized watermark is less than length of cover signal perform embedding to produce Normalized Speech watermark signal.

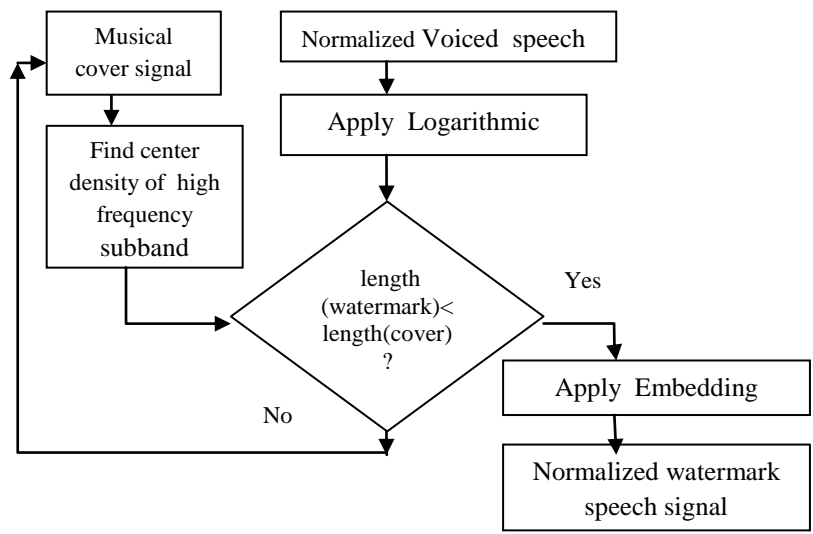

Figure 4: Normalized speech watermark

## 5. EXPERIMENTAL SETUP AND RESULTS

The corresponding code of the proposed technique has being implemented in MATLAB. The secret code such as password FIVE has been given as real time input. Female speaker has recorded the password with sampling frequency(FS) 8000 Hz and duration 3 seconds  By listening the input waveform as well as by voice activity detection techniques the presence of password in the signal has been confirmed. Since the signal is non stationary and changes with time frame by frame analysis is done so that no information is lost. Based on a visual evaluation on the graphically representation as shown in figure and corresponding computed values voice activity detector classify the input signal as voiced speech signal. The short time energy is measured as 0.0116 and zero crossing  rate as 0.074979.

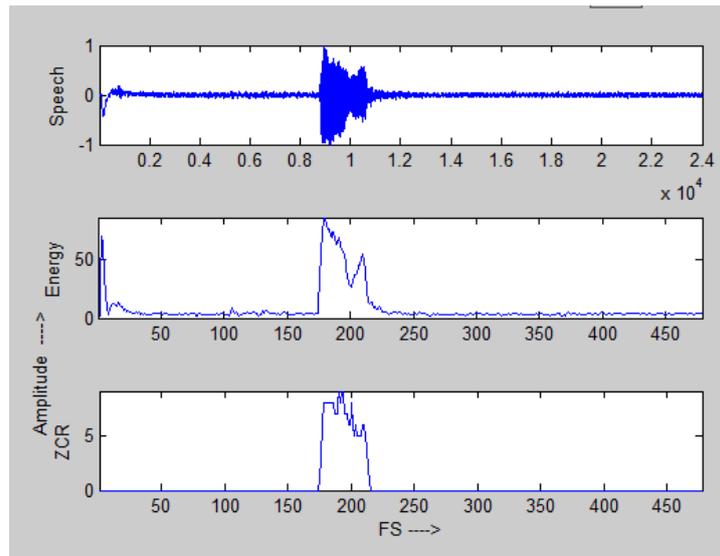

Figure 5: Signal with password Five, energy and zero crossing rate plot.

The energy represents the variations in amplitude and maximum amplitude of the voiced speech signal was found to be 0.9853 i.e. 9853 Hz. Once the speech signal is classified as voiced speech, to remove noise wiener filter is applied. Using cepstral based method cepstral peak of the signal is estimated and it is observed as 200. For further processing we have considered voiced speech frame. The frequency of the voiced frame is processed in voice normalization phase and the MATLAB code used for normalization is

Y= max(max(abs(speech))); % speech – cropped voiced speech
N = 1/log(1+(Y));          % Ratio for Normalization
FreqNorm = N* log(1+abs(speech));
Listening to the normalized voiced speech it was found satisfactory and it is given as watermark to speech watermarking module. Logarithmic /exponential operation is applied to it to make it more secure. The non-voiced musical sequence with frequency sampling 22050 Hz is considered as cover signal. The length of the watermark is found to be 2126 and that of cover signal as 70641. In the center density of highest frequency subband of host signal the normalized watermark is embedded using frequency masking approach and the outcome is normalized speech watermarked signal. Demonstration tests were performed and the results were satisfactory.

## 6. CONCLUSION AND DISCUSSION

The proposed algorithm aims at providing robust secured technique for password transmission. Using logarithmic-exponential method the security level of watermarking could be enhanced. Researchers have proposed many techniques to improve security level of speech watermarking but little attention has been given to speaker identity who is sending the data to withstand the security attack. To take additional measure the normalization technique has been used in conjunction with speech watermarking technique.

We found that normalization and embedding strategy was satisfactory since the speaker identity cannot be revealed, watermark that is information carried by the host signal is imperceptible.

This algorithm can be improved considering issues close to the robustness of the methods against various security attacks